\documentclass[paper=a4, fontsize=11pt]{scrartcl}
\usepackage[T1]{fontenc}
\usepackage{fourier}

\usepackage[english]{babel}															% English language/hyphenation
\usepackage[protrusion=true,expansion=true]{microtype}	
\usepackage{amsmath,amsfonts,amsthm} % Math packages
\usepackage[pdftex]{graphicx}	
\usepackage{url}

\usepackage{sectsty}
\allsectionsfont{\centering \normalfont\scshape}

\usepackage{fancyhdr}
\numberwithin{equation}{section}		% Equationnumbering: section.eq#
\numberwithin{figure}{section}			% Figurenumbering: section.fig#
\numberwithin{table}{section}				% Tablenumbering: section.tab#

\newcommand{\horrule}[1]{\rule{\linewidth}{#1}} 	% Horizontal rule

\title{
		\usefont{OT1}{bch}{b}{n}
		\normalfont \normalsize \textsc{Department of Physics, University of California, Davis} \\ [25pt]
		\horrule{0.5pt} \\[0.4cm]
		\huge Gender-grade-gap zeroed out under a specific intro-physics assessment regime \\
		\horrule{2pt} \\[0.5cm]
}
\author{
		\normalfont 								\normalsize
        David J. Webb and Wendell H. Potter\footnote{Deceased 8 January 2017}\\[-3pt]		\normalsize
        \today
}
\date{}

\begin{document}
\maketitle
\section{Introduction}
Unhappiness with issues involved in grading, coupled with a desire to offer classes where everyone succeeded, led us to offer several classes of a large-enrollment introductory physics course that took a small step toward mastery learning.  These attempts to improve instruction in these courses involved three different instructors.  The classes were offered about 5 years ago but I (David Webb) retired almost immediately after those course offerings and my colleague Wendell Potter died less than a year later so I have only recently returned to these data.  This report is written from my perspective.

In the last decade or so, when I have done test runs of significant systemic changes in introductory physics courses I have checked whether the particular changes we made discriminate, in a way that is measurable by grades, against any large demographic groups in the course.  The concern is mainly that people who already face discrimination are not further hurt by our changes in these courses.  The American Physical Society (APS) has identified several groups of people who are represented in the field of physics in the United States at smaller percentages than they are in the larger US population.  The APS groups these people in terms of their gender (female physicists\footnote{Both the APS and our university data have had a background assumption of gender as a binary category.}) and/or their racial/ethnic ancestry (physicists with african, native american, hispanic, or pacific island heritage).  There are usually a large enough group of female students to allow statistical studies.  However, we must usually gather students with these racial/ethnic ancestries together into a single group to give us the statistical power to do these checks.  We use the general term Underrepresented Minority (URM) students for this group of students.  Finally, we sometimes have enough data for studies of students in the intersections of these groups.

In comparing the grades from these mastery trials to those from our usual class experiences we noticed that the gender gap we find in grades in our usual classes may just be an artifact of the organization/assessment/grading methods in those classes.  This is a report on the trial classes and some results.  First, I’ll discuss how course grades were usually determined and our unhappiness with this grading.  Then I’ll discuss the systemic changes in organization/assessments/grading in the trial run classes.  Finally, I’ll present some data on the gender-grade-gap, comparing the modified classes with classes where organization/assessments/grading were more standard.

\section{Background on CLASP courses}
The active-learning based CLASP (Collaborative Learning through Active Sense-making in Physics) courses used in this report have been described in detail in previous work \cite{Potter2014Sixteen}.  Basically, CLASP features an explicit focus on understanding models (including words, graphs, equations, etc.) and using these models in qualitative and quantitative analysis of real-world physical, chemical, and biological situations.  In a usual 10-week term the course includes a single 80 minute lecture meeting per week coupled with two 140-minute discussion/laboratory (DL) sections per week.  The active learning elements of the course are carried out in these studio-type DL sections of 25-35 students.  The DL sections include small group activities where students work together to understand a model and practice applying it before engaging in whole-class discussions.  There are three courses, CLASP A, CLASP B and CLASP C, making up this one-year introductory-physics series.  The courses are meant to be taken in sequence and cover essentially all of the introductory physics in a standard series for biological science students.

\section{Categorical grading in CLASP}
Instructors in most CLASP classes grade exams using a numerical grading system \cite{Webb2020Grading} that directly links every graded item (be it an exam question, or the exam itself) to an absolute grade scale so that students can always understand how their performance on a given question related to the expectations of the course instructors.  Specifically, each possible letter grade (A+, A, A-, B+, etc.) is represented by a range of numbers on the grade scale that an instructor uses.  There are two main grade scales used by CLASP instructors, a 4-point-based scale (CLASP4) and a 10-point-based scale (CLASP10).  These two scales are discussed in detail in reference \cite{Webb2020Grading}.  In practice, the graders use a scoring method called ``grading by response category'' (GRC) \cite{Paul2014Grading} in which a grader would categorize student responses by their most significant error, and the instructor would assign the same score and written feedback to all students who made that error.  This type of scoring cannot be considered a rubric because the categories are made after looking at student responses, but are otherwise similar to holistic rubrics in that scoring is subjective (requires judgment as the answer is not simply correct or incorrect) and that a single score and feedback is given for each exam problem.

\section{Unhappiness with this grading}
Regarding GRC, my colleague Wendell Potter often pointed out that much of the time a grader spent trying to carefully distinguish the physics value of various students’ answers was wasted.  In his view there is a very basic division into two groups, either a student ``got it'' or they ``didn’t get it'' with the dividing line between these two groups likely somewhere in the B$-$ to B$+$ grade range.  This basic division into two groups could be done by a grader with minimal cost in time (just reading the student’s answer) for the vast majority of students.  And, finally, that parsing the various answers from students who ``didn’t get it'' was a waste of the grader’s time.  Basically, a student making a major physics error or many small errors or omitting an extremely important issue from their discussion ``didn’t get it'' and those students giving answers with even less correct physics content than this also ``didn’t get it''.  Wendell considered it a waste of a grader's time and also a hopeless task to attempt to divide unacceptable answers into categories and then decide which categories showed understanding that was satisfactory (the C's), which should be labeled poor (the D's), and which should be labeled failing (the F's).

In a recent paper \cite{Webb2020Grading} we presented data suggesting that Wendell’s vocal worries about parsing the grades given to the answers of students who ``didn’t get it'' were, perhaps, justifiable.  We showed\cite{Webb2020Grading} that D and F grades (and also C and D grades) seem quite fungible.  In comparing the two main grade scales, CLASP4 and CLASP10, we noted that instructors seemed to shift about 15\% of the total grade weight from from C and D grades, given under CLASP4, down to F grades when the instructor used the CLASP10 grade scale.  The fungibility of C, D, and F grades strongly suggests that the answers from students who ``didn’t get it'' were not easily placed onto an absolute grade scale by \textbf{any} of the seven instructors who used both grade scales at various times.

\section{Mastery teaching and standards-based grading}
To me, offering a class devoted to mastery means, at a minimum, giving students whatever time they need to reach mastery of each of the various topics in the course and, importantly, assessing their work as necessary to provide them with a gauge of that mastery.  In CLASP these assessments most often come during DL where they are short verbal assessments but can also come during an active-learning lecture and, of course, could include the timed exams that students take.  Implicit in a mastery class is the instructor's confidence that every student can succeed in mastering every topic.

The usual way of providing the gauge for a student's work is called ``standards-based grading''.  In standards-based grading the goals that the instructor has for a student are written in a set of standards and the standards that each student should meet are known to the students during their work.  The student's work is judged against each standard using a grade scale including only a few levels like ``below basic'', ``basic'', ``proficient'', and ``advanced''.  The grades ``proficient'' and ``advanced'' would include all of the students who ``got it'' according to Wendell.  A description of some attempts at a completely mastery physics class using standards-based grading is given by Beatty \cite{Beatty2013SBG}.  The trial classes described in the present report did not aim at the full implementation that Beatty tried because those seemed too difficult for us to implement in our large classes.  Nevertheless, we made a few important changes toward mastery classes in our trial runs.

\section{Experiments in grading}
\subsection{Classes during summer terms}

Wendell and I decided to try out a new grading scheme in a summer quarter in 2015.  I taught two classes, one CLASP A and one CLASP B, that quarter and both became trial runs of this teaching method.  CLASP courses are condensed by a factor of about two during a summer session so that almost all of the course fits into 5 weeks instead of a normal 10 week quarter.  In the summer there are two 75 minute lecture times per week and four 140 minute DL meetings per week.  We made three major changes to these courses.

The first change was toward mastery learning.  Giving students many chances to demonstrate their mastery didn't seem feasible but we needed to give students at least a second time to demonstrate mastery of each topic.  We divided the topics into five parts and students took four quizzes (on the first four topics) in their DL's during the quarter.  About a week after each of the four quizzes the students were given a chance to take a new quiz, a retake given during lecture time, covering the same material as the regular quiz they had taken the week before.  If the student received a higher score on that retake quiz then the retake score replaced their first score.  If the student didn't get a higher score then the retake score was averaged with the original score and the average was used as the measure of their mastery of that topic\footnote{As a practical matter, students who missed a quiz took the retake quiz as a makeup but did not have their own second chance.}.

The second change was to the lecture time.  I ran the lecture time in a flipped-course style by offering online lectures (complete with conceptual questions) for students to view before the lecture time.\footnote{Our department had recently finished recording and putting lectures onto YouTube so that we could easily offer these lectures online.}  Then I used the live-lecture times as a kind of office-hour/practice-session where students worked on old quiz problems with my help.  Of course one of the two lectures each week had to have time saved at the end for a retake quiz for students who chose to try to improve their grade.

The third change we made was in the grading.  As pointed out above we decided that a standards-based grade regime was too difficult to implement.  But, as a step toward standards-based grading, Wendell and I decided that students whose answers were essentially perfect (no physics errors but maybe a small math error) received a grade = 1.0.  These students' answers would have shown them to be ``advanced'' in their work on the subject of the quiz (greater than or maybe equal to A$-$ using our usual GRC grading).  Students making a minor physics error or with a minor part of their explanation missing were given a grade = 0.67.  These students' answers would have shown them to be ``proficient'' in their work on the subject of the quiz (between a B and an A$-$ or A using our usual GRC grading).  All other answers would receive a grade = 0.\footnote{To choose these scores (and the letter grade cutoffs described later), we re-scored students from a previous offering of CLASP A to find grade cutoffs that would produce approximately the same course grade distribution as the original GRC grading had for that previous class.  We re-scored all original grades from a low B+ through middle A- with the ``proficient'', 0.67, grade and higher original grades with the ``advanced'', 1.0, grade in this re-scoring.}  Our feeling about the answers showing ``proficiency'' is that they were close enough to correct that it seems likely that the student either forgot to deal with some small issue, or didn’t notice some small issue and that the student likely would have been in the ``advanced'' group with only a short comment from the instructor to guide them.

Students took a final exam in each of these courses but there was no time during the term to allow a retake on the final.  The final exam was graded on the same scale as the quizzes (every answer received either 0, or 0.67, or 1) and the scores were averaged using a weighted average to give a total score less than or equal to 1.0.  Course grades were given in a fairly standard way by choosing cutoffs for each grade.  The A+,A,A- range lower cutoff was 0.70, the B+,B,B- range lower limit was 0.49, and the C+,C,C- range lower limit was 0.27 with all other grade cutoffs appropriately placed.  So students had to average better than ``proficient'' to receive a grade higher than B+.

\subsection{Classes during Fall and Winter terms}
After the summer session trials I taught another class (a CLASP C class) that used retake quizzes and we recruited two more CLASP A instructors to try offering retake quizzes in their classes\footnote{Most instructors did not want to have to give up their lecture time to online lectures and/or did not want to have to offer extra quizzes and/or did not want to use such a coarse grading method}.  These three classes all used online lectures and gave retake quizzes but there were two distinctly different quiz-offering schemes.  Importantly, these two distinct quiz-offering schemes had distinctly different results regarding the gender gap.

In my CLASP C class and one of the CLASP A classes a quiz was given during lecture every two weeks and the retake quizzes were given in lecture during the weeks when no new quiz was given.  The online lectures allowed these instructors to spend the rest of their official weekly lecture time (45-50 minutes) helping students work on old quiz problems as practice (as was done during the summer lectures).  So, in these classes students ended up with grades on four quizzes and never took more than one quiz during any particular lecture time.  For these classes the retake grade was substituted for the original grade if it was higher and was averaged with the original grade if it was not higher and retakes were not allowed on the final exams.  This regime roughly replicated the summer assessment regime.

The other CLASP A instructor was interested in testing out a few new ideas.  First, the idea that the act of taking an exam led to learning.  This idea led this instructor to want to give many quizzes to increase learning.  Second, the idea that because we want students to be creative, the quizzes themselves should ask the students to be creative.  This idea led to some exams where I had difficulty figuring out what the instructor was asking and how I should approach a solution.  Many of this instructor's quiz questions were very different from any I had ever seen in these courses.  In addition I think that they could be considered quite difficult questions.  Both these ideas went very well with allowing retakes so this instructor gave an original quiz during lecture time every week (except the first week and the last week) and then, during the same lecture time, a retake of the previous week's quiz.  Specifically, during a typical lecture time in this class the instructor i) gave a 30 minute quiz on new material, ii) used 15 minutes to review the quiz the students just took, and then iii) gave a 30 minute retake quiz.  So, in this class students ended up with grades on eight quizzes and, in addition, almost always took two quizzes during a lecture meeting.  For this class the highest of the two quiz grades was always used as the official quiz grade.  As is often the case when changes are made, this instructor did not make any measurements testing whether these particular exam changes led to any increases in learning.  Needless to say, this class is likely to be anomalous so I will separate out these data for special consideration.

\section{Data and analysis methods}
In comparing gender-grade-gap in the mastery CLASP classes with that of the more canonical CLASP classes we will use data from about three years (winter of 2013 through winter of 2016) that includes all 67 CLASP classes (A, B, and C parts are all included) offered in that time period. The classes changed very little during those years.  All together these 67 classes included 17,205 grades given to students.  To address the student response to these classes we also gave an anonymous survey.

The university administration supplied us with the self-identified (binary) gender of each student\footnote{Problems with defining gender as a binary are well known and important but this report on trial classes from a few years ago is not able to address these issues in any statistical way.}.  There were 10,937 female students, 6,266 male, and 2 students unidentified as either male or female.  We will compare the mastery classes with the usual GRC grading classes.  I should note that one of the instructors offering a mastery class had previously taught courses using our online lectures so that we will be able to see that offering lectures online did not, by itself, diminish the gender gap.  Because there were two distinct ways to offer retake exams we will separately analyze their effects.

We have shown\cite{Webb2020Grading} that the class grade distributions may be significantly different depending both on the instructor and the grade scale used.  In order to minimize these effects in our examination of gender gaps, we follow Ref. \cite{Salehi2019Gaps} and use the relative grade of a student rather than their absolute grade.  We normalize the absolute grade on a class-by-class basis to define Zgrade so that the Zgrade distribution has the same average, 0, and standard deviation, 1, for each class.

Because students are grouped into classes we use Hierarchical Linear Modeling (HLM) and include two levels, a student level and a class level.  The categorical variable ``Female'' is a student level variable and the categorical variable ``Mastery'' denotes a mastery class and so is a class level variable.  Although we are using HLM in our analyses the results would be almost exactly the same if we had done simple regression or just t-tests.

\section{Results}
\subsection{Student and TA responses to the mastery classes}
We surveyed our students anonymously after offering these classes.  Students responded to statements using a 5-point Likert scale (Strongly agree, Agree, Neutral, Disagree, Strongly Disagree).  The two statements whose responses were most important to us were 1) ``I would choose a CLASP class that had ``acceptable/unacceptable'' grading of exams \textbf{without being offered retake exams}.'' and 2) ``I would choose a CLASP class that had ``acceptable/unacceptable'' grading of exams \textbf{if I were offered retake exams}.''  Only 36\% of the students were either neutral to or supportive of the first statement but 91\% of the students were neutral to or supportive of the second statement.  So students were mostly happy with this grading method but only if the class was aimed toward mastery of the material by offering retake exams.

The Teaching Assistants (TAs) in the course ran the DL's and did the grading of quizzes and they had one main complaint about this kind of grading.  They uniformly wanted there to be one more category between ``proficient'' and ``unacceptable''.  Their reasoning was that they were unhappy giving a student's answer 0 points if it showed considerable knowledge of the subject.  I understood their issue but did not change the grade scale because those students who showed considerable understanding still were giving answers that did not, in my view, show ``proficiency''.

\subsection{The gender gap is gone}

\subsubsection{Mastery classes giving only one quiz during any lecture time}
For the classes where students took at most one quiz per week in lecture we first note that women and men came into the classes with similar GPA's as shown in Table \ref{tab1} so we will just compute the gender gap without controlling for demonstrated academic successes.  We compare the gender gap for our mastery classes with the gender gap for classes with the usual assessment/grading regime.  Using HLM to model the gender gap directly at first as:
\begin{align} 
	\begin{split}
	CourseGrade 	&= b_{0}\\
					&+ b_{Female}(Female)
	\end{split}					
\end{align}
where Female is a categorical variable equal to 1 if the student self-identified as Female and equal to 0 if the student self-identified as Male.  We fit mastery courses and the canonical courses separately so that we have a gender gap, given by $b_{Female}$, for each.  At this point we do not control for any other characteristics of the students.  In Table \ref{tab2} we give the results for these two separate gender gaps.  The usual gender gap shows that the grades of female students were almost a quarter of a standard deviation below those of male students for canonical grading in CLASP courses.  On the other hand, for the mastery classes female students had slightly higher grades than male students but not significantly higher.  So our conclusion is that there was no gender gap in the one-quiz-per-lecture group of CLASP courses offered as mastery courses even though there was a significant gender gap in the usual CLASP courses.  These are the main points of this report and neither of these conclusions would change if we had controlled for the students' incoming GPA's.\footnote{One instructor had used the flipped-class format with the usual grading methods.  This class had a gender gap of $-0.22 \pm 0.07$.  So, the gender gap seems more related to the assessment/grading scheme that to the flipped-class issue.}

\begin{table}
\caption{Incoming GPA's for both female students and male in groups of courses considered in this report.  The groups include 1) the group of mastery courses that offered one quiz per lecture time (M1Q), 2) the single mastery course that offered both a quiz and a retake quiz in most lecture times (M2Q), and 3) all courses from Winter 2013 through Winter 2016 that did not offer any retake quizzes (Usual).  The t-statistics and P-values are for two-tailed t-tests comparing the GPA's and suggest that female student's incoming GPA's are statistically indistinguishable from male student's incoming GPA's in each group.}

\label{tab1}
\begin{center}
\begin{tabular}{c c c c c c c}
\textbf{Group} & \textbf{N Female} & \textbf{GPA Female} & \textbf{N Male} & \textbf{GPA Male} & \textbf{t-statistic} & 
\textbf{P-value} \\ 
 \hline
Usual & 9,785 & 3.09 & 5,480 & 3.08 & 1.16 & 0.245 \\
M1Q & 400 & 3.07 & 210 & 3.01 & 1.41 & 0.158 \\
M2Q & 169 & 3.16 & 85 & 3.10 & 0.977 & 0.330 \\
\end{tabular}
\end{center}
\end{table}

\begin{table}
\caption{Gender gap (negative gap means female students are given lower grades) for the same groups from Table \ref{tab1}.  N is the number of grades given to students in each group.}
\label{tab2}
\begin{center}
\begin{tabular}{c c c c c }
\textbf{Group} & \textbf{N} & \textbf{GenderGap} & \textbf{Stand. Error} & \textbf{P-value} \\ 
 \hline
Usual & 16,246 & -0.223 & 0.016 & $<10^{-3}$ \\
M1Q & 636 & 0.025 & 0.083 & 0.766 \\
M2Q & 321 & -0.21 & 0.12 & 0.072 \\
\end{tabular}
\end{center}
\end{table}

\subsubsection{Two quizzes a day}
The course that offered two quizzes in most of the lectures had a quite different result.  Table \ref{tab2} shows that the gender gap in this course was similar to that in CLASP courses with the usual kind of grading.  Again, Table \ref{tab1} suggests that this gender gap in grades was not due to a gender gap in demonstrated academic skills.

\subsection{Grade gaps of underrepresented groups}
The university has also supplied us with the self-identified ethnicity of these students so we can check to see whether changes in assessments toward mastery also have effects on grade gaps sometimes seen\cite{Webb2017Concepts}\cite{Salehi2019Gaps} between students from underrepresented racial/ethnic groups and their peers in introductory physics\footnote{In keeping with previous work, we only include US citizens and non-citizens with permanent resident status in the URM and nonURM groups.}.  There were 3,173 students in the URM category and 13,729 students nonURM and 303 of unknown or unstated heritage.  Again we use HLM to model this gap directly as:
\begin{align} 
	\begin{split}
	CourseGrade 	&= b_{0}\\
					&+ b_{URM}(URM)
	\end{split}					
\end{align}
where URM is a categorical variable equal to 1 if the student self-identified as belonging to a racial/ethnic group underrepresented in physics and equal to 0 if the student did not so identify.

The results of this analysis are shown in Table \ref{tab3}.  These URM grade gaps are statistically indistinguishable so these changes in the assessment scheme toward mastery learning have negligible effect on URM grade gaps\footnote{These URM grade gaps are likely due to systemic racism in a way that is noted in reference \cite{Webb2017Concepts} as well as later in this report.}.

Finally, the group of students who are members of both categories, female and URM, experience an average grade gap that is larger than expected for either group but a little smaller than the sum of the two independent grade gaps.  That is, in the usual classes this inter-sectional group ($N=1,941$) experiences a grade gap (when compared to the rest of the class) of $-0.478 \pm 0.024$.  Conversely, in the M1Q trial classes, this inter-sectional group ($N=96$) experiences a grade gap (when compared to the rest of the class) of $-0.26 \pm 0.11$.

\begin{table}
\caption{Racial/Ethnic grade gap (negative gap means students from underrepresented groups are given lower grades) for the same groups from Table \ref{tab1}.}
\label{tab3}
\begin{center}
\begin{tabular}{c c c c c }
\textbf{Group} & \textbf{N} & \textbf{URMGap} & \textbf{Stand. Error} & \textbf{P-value} \\ 
 \hline
Usual & 15,313 & -0.375 & 0.020 & $<10^{-3}$ \\
M1Q & 603 & -0.328 & 0.093 & $<10^{-3}$ \\
M2Q & 293 & -0.56 & 0.14 & $<10^{-3}$ \\
\end{tabular}
\end{center}
\end{table}

\section{Discussion}

We did these original trial runs in an effort to allow all of our students to maximize their learning and all of our students to demonstrate their mastery of the various topics of the course.  We knew that we were attempting systemic changes in the courses but at no point in the planning discussions did I have any idea that we would end up discussing a likely case of systemic sexism.  Nevertheless, that is my conclusion and one that Wendell recognized also.

These trial classes were designed to help us learn about how we might offer a mastery class and so they don't give us much help in identifying a mechanism for how it is that they seem to have changed a class with a measurable gender gap into one that is essentially gender neutral.  The trial classes can be viewed as a usual class coupled together with an independent study class that runs in parallel with the usual class.  The course carries on with standard lectures and discussion/labs as usual the entire term but, after the first quiz, there are groups of students who are simultaneously studying independently for their retake quiz.  We have essentially no information on what students are doing during this independent study part of the course and so have little information helping us understand exactly how the class changes have changed the students.

Even though we have no data speaking to how these trial classes affect the students, we can suggest some possibilities:

I) The explicit focus on mastery may lead students to both understand and value mastery more than they would have.  In education literature there are suggestions\cite{Brookhart1997Assessment}\cite{Dweck2007Gender}\cite{Kelly2016SocialCogn} of connections between mastery orientation, ideas of intelligence being malleable (growth mindset), intrinsic motivation toward learning, self-efficacy, and assessment types.

II) The fact that the instructor explicitly values and supports student learning (even if they have not shown mastery on their first exam) may be important for reasons other than I) above.

III) The feedback and extra (independent) studying may simply lead to more clarity in thinking, and deeper learning, about the topic in question.

IV) It may be important, for some other reason than those listed above, for students to cycle back through something that they have already been tested on.

V) It may be that simply having time pass after first study of a topic leads to greater and deeper understanding even in the absence of any cycling back through the material.

Possibilities I) and II) (and maybe even III) and IV)) could also be used to help explain the difference between a rather relaxed mastery class (M1Q) whose students complete at most one quiz per lecture time and a rather rugged mastery class (M2Q) whose students take two nearly sequential quizzes in most lectures.

\begin{table}
\caption{Groups whose demographic grade gaps are shown in Fig. \ref{Fig1}, the measurements for which those gaps were defined, and the studies from which the data were acquired.}
\label{tab4}
\begin{center}
\begin{tabular}{| p{16mm} | p{55mm} | p{25mm} | c |}
\hline
\textbf{GroupID} & \textbf{Group} & \textbf{Measurement} & \textbf{Ref.} \\ 
 \hline  \hline
HSEC & Introductory physics classes for \newline engineers and physical science \newline majors at a highly selective east \newline coast university & Final exams & Salehi et. al. \cite{Salehi2019Gaps} \\ \hline
HSWC & Introductory physics classes for \newline engineers and physical science \newline majors at a highly selective west \newline coast university & Final exams & Salehi et. al. \cite{Salehi2019Gaps} \\ \hline
PM & Introductory physics classes for \newline engineers and physical science \newline majors at a public university in the \newline middle of the country & Final exams & Salehi et. al. \cite{Salehi2019Gaps} \\ \hline
CLASP \newline Usual & CLASP classes, usual GRC grading & Course Grades & This report \\ \hline
CLASP \newline M1Q & CLASP classes, mastery grading, one quiz per lecture & Course Grades &  This report \\ \hline
CLASP \newline M2Q & CLASP classes, mastery grading, two quizzes per lecture & Course Grades & This report \\ \hline
UCD \newline Engin. & Three introductory UCD physics classes for engineers and physical science majors (usual organization of topics) & Final exams \newline (same exam as \newline class below) & Webb \cite{Webb2017Concepts} \\ \hline
UCD \newline Engin. \newline Concepts \newline First & One introductory UCD physics \newline class for engineers and physical \newline science majors (concepts taught \newline before any complicated \newline calculations) & Final exams \newline (same exam as \newline classes above) & Webb \cite{Webb2017Concepts} \\ \hline
\end{tabular}
\end{center}
\end{table}

\begin{figure*}[h]
\includegraphics[trim=1.3cm 1.6cm 1.6cm 1.0cm,clip=true,width=\linewidth]{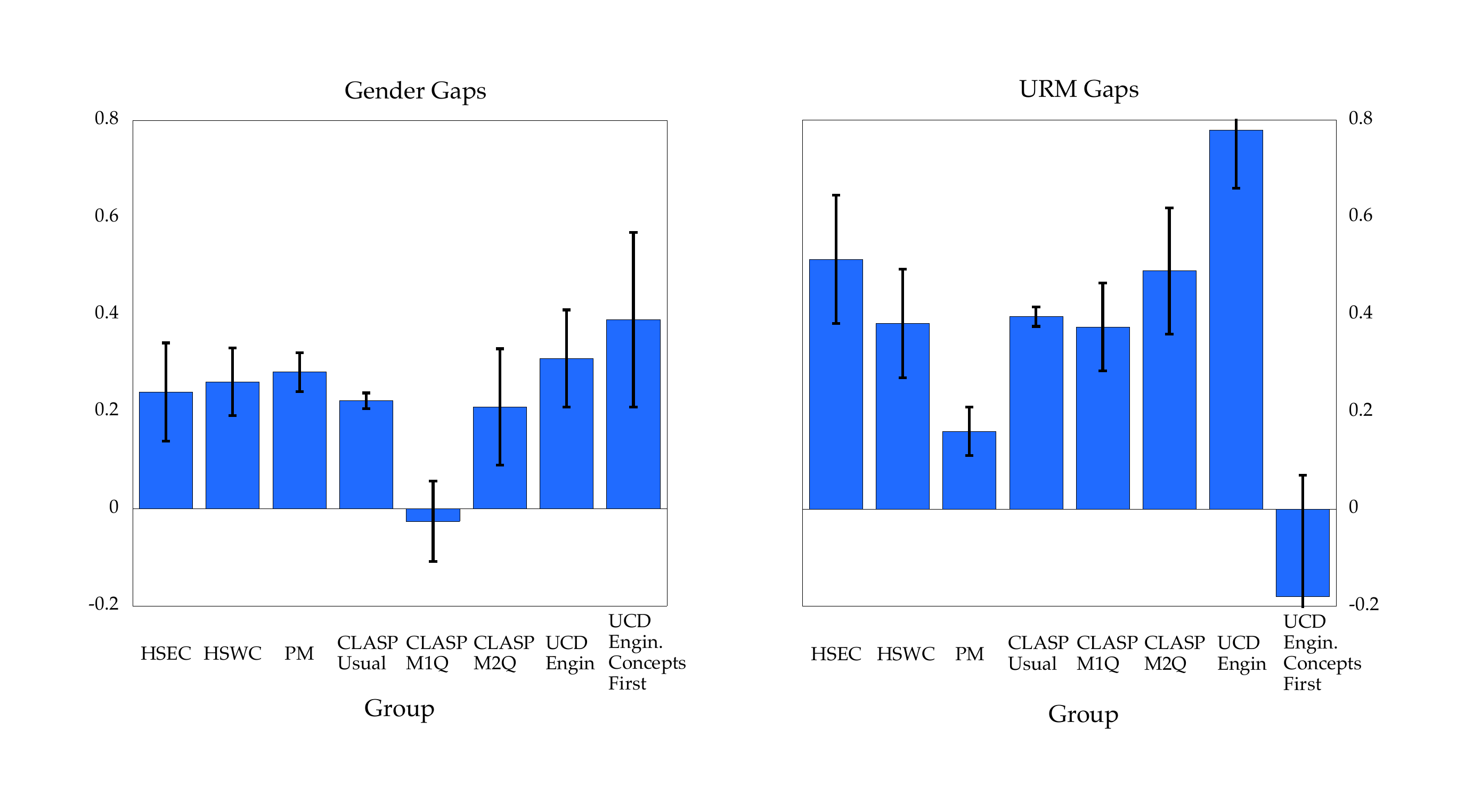}
\caption{Demographic gaps from several groups defined in Table \ref{tab4}.  The grade gaps are all in units of standard deviations of the distribution.  The groups HSEC, HSWC, and PM are from Ref. \cite{Salehi2019Gaps} and show grade gaps on final exams.  The groups CLASP Usual, CLASP M1Q, and CLASP M2Q are from the present report and show course grade gaps (following Salehi I am making gaps positive if women or students from underrepresented groups had lower grades).  The groups UCD Engin. and UCD Engin. Concepts First are an analyis of the data from Ref. \cite{Webb2017Concepts} and show grade gaps on final exams.}
\label{Fig1}
\end{figure*}

As a broader look at these issues, I want to replot data from this paper and from an earlier study \cite{Webb2017Concepts} and compare with a recent paper \cite{Salehi2019Gaps} by Salehi et. al.  Figure \ref{Fig1} shows demographic gaps for several groups of classes that are defined in Table \ref{tab4}.  Note that I am following Salehi et. al. now in using \textbf{positive gaps} when the affected group (either women or students from underrepresented groups) have \textbf{lower grades}.  We have previously noted \cite{Webb2020Grading} that grades in CLASP are almost completely due to exam scores\footnote{I estimate that differences between the gaps in CLASP course grades and the gaps in CLASP exam grades are less than one-third of the CLASP Usual standard error in Fig. \ref{Fig1}.} so I would argue that the CLASP course grade gaps can be reasonably compared to the final exam grade gaps studied by others.  There are three points that I would make about this figure:

1) Gender gaps are present and of similar magnitude in these classes when they are offered in their usual way.  URM gaps are also present in the usual classes but seem more variable.

2) One may be able to zero-out either gap by making a systemic change that improves the course for all students.  For the gender gap we see this in the mastery classes and for the URM gap we see it in the concepts-first course.  If either gap can be zeroed out by changing the course then one should probably not assume that introductory physics courses and/or their exams/grades are demographically neutral.  Instead, one should probably allow for the possibility that each introductory course is institutionally racist or sexist (or both) in its own way and by its own amount.

3) A course improvement that zeros-out the gender gap may be quite different from a course improvement that zeros-out the URM gap.

Points 2) and 3) suggest, to me, that teachers should sometimes be willing to make rather large systemic changes aimed at improving learning as long as they can study the results of those changes to decide who was helped, who was hurt, and who was unaffected.

%%% End document

\begin{thebibliography}{15}

\bibitem{Potter2014Sixteen}
Wendell Potter, David Webb, Cassandra Paul, Emily West, Mark Bowen, Brenda Weiss, Lawrence Coleman, and Charles De Leone, ``Sixteen years of collaborative learning through active sense-making in physics (CLASP) at UC Davis'', Am. J. Phys.,
\textbf{2}, 153--163 (2014).


\bibitem{Webb2020Grading}
David J. Webb, Cassandra A. Paul, and Mary K. Chessey, ``Relative impacts of different grade-scales on student success in introductory physics'', Phys. Rev. Phys. Educ. Res., \textbf{16}, 020114, 1-17 (2020).


\bibitem{Paul2014Grading}
Cassandra Paul, Wendell Potter, and Brenda Weiss, ``Grading by Response Category: A simple method for providing students with meaningful feedback on exams in large courses'', Phys. Teach., \textbf{52}, 485-488 (2014).


\bibitem{Beatty2013SBG}
Ian D. Beatty, ``Standards-based grading in introductory university physics'', J. Scholar. Teach. Learn., \textbf{13}, 1-22 (2013).


\bibitem{Salehi2019Gaps}
Shima Salehi, Eric Burkholder, G. Peter Lepage, Steven Pollock, and Carl Wieman, ``Demographic gaps or preparation gaps?: The large impact of incoming preparation on performance of students in introductory physics'', Phys. Rev. Phys. Educ. Res., \textbf{15}, 020114, 1-17 (2019).


\bibitem{Webb2017Concepts}
David J. Webb, ``Concepts first: A course with improved educational outcomes and parity for underrepresented minority groups'', Amer. J. Phys., \textbf{85}, 628-632 (2019).



\bibitem{Brookhart1997Assessment}
Susan M. Brookhart, ``A theoretical framework for the role of classroom assessment in motivating student effort and achievement'', Appl. Meas. Educ., \textbf{10}, 161-180 (1997).


\bibitem{Dweck2007Gender}
C. S. Dweck, ``Is math a gift? Beliefs that put females at risk.'', \textit{Why aren’t more women in science?  Top researchers debate the evidence}, Ed. S. J. Ceci and W. M. Williams, American Psychological Association, Washington, DC, 47– 55, (2007). 

\bibitem{Kelly2016SocialCogn}
Angela M. Kelly, ``Social cognitive perspective of gender disparities in undergraduate physics'', Phys. Rev. Phys. Educ. Res., \textbf{12}, 020116, 1-13 (2016).


\end{thebibliography}
\end{document}